# Interactions between Josephson and pancake vortices revealed by second harmonic Hall measurements

Mauro Tesei and Lesley F Cohen
*The Blackett Laboratory, Physics Department, Imperial College London, SW7 2AZ, UK*

The dynamic interactions between Josephson and pancake vortices in highly anisotropic superconductors are probed by shaking the former vortices with an ac magnetic field and measuring the second harmonic Hall response of the latter vortices with the sensor positioned above the sample. We demonstrate the efficiency of the proposed measurement method that aims to probe the coupling between the two vortex species. New experimental data obtained on $Bi_2Sr_2CaCu_2O_8$ single crystals reveal interesting features and the possibility to determine the phase diagram of vortex matter in such superconductors.

## I. Introduction

Magnetic field penetrates a homogeneous high temperature superconductor (HTS) in the same way it penetrates a type-II superconductor, i.e. in terms of Abrikosov vortices. Furthermore, if the sample is disorder-free (i.e. no pinning centres) the vortices arrange themselves in a regular triangular lattice. When the HTS is highly anisotropic like $Bi_2Sr_2CaCu_2O_8$ (BSCCO) which is formed by weakly coupled parallel superconducting *ab* planes, the penetration of the magnetic field strongly depends on its orientation with respect to the *ab* planes. The anisotropy is characterised by the ratio of the perpendicular penetration depth $\lambda_Z$ and the in-plane penetration depth $\lambda_{ab}$. Typically, for BSCCO single crystals $\lambda_Z/\lambda_{ab} \sim 500$ [1]. When the magnetic field is perpendicular to the *ab* planes only stacks of two-dimensional "pancake" vortices (PVs) [2] are created within the *ab* planes, which contain the supercurrents as well as the normal core of a PV. In this case PVs arrange themselves in a regular triangular Abrikosov-like lattice. When the magnetic field is parallel to the *ab* planes elongated Josephson vortices (JVs) are formed with the normal core of a JV lying in the non-superconducting region between two neighbouring *ab* planes [1]. In the case of tilted magnetic field, PVs and JVs coexist and the regular PV lattice is modified due to the attractive interaction with JVs [3], giving rise to a rich vortex phase diagram [1] which depends on both the in-plane and perpendicular components of the applied magnetic field. Although there has been a great interest in the interplay between the two sets of vortices with considerable potential applications, envisaging a new type of vortex logic based on the manipulation of single PVs by controlling the JV lattice [4], the dynamical interaction has essentially been studied through the dragging of PVs by JVs in the so-called lensing and ratchet experiments [5-8].

In this paper we show new experimental results on the dynamic interactions between the two vortex sublattices using a scanning Hall probe combined with an ac magnetic field technique [9]. The observed variations in the strength of the JV-PV interaction reveal a crossover consistent with the expected phase diagram for vortex matter in BSCCO single crystals [10] and definitively show the potential of the measurement method in probing the various vortex states in anisotropic superconductors.

## II. The experiment

PVs are introduced in a BSCCO single crystal by field cooling the sample in a perpendicular magnetic field $H_Z$ produced by a solenoid coil. The perpendicular magnetisation associated with PVs is monitored using an InSb Hall sensor (Te-doped 3.5μm thick InSb on GaAs wafer with mobility $\mu(300K)=5.8m^2/Vs$) of active area $100 \times 100 \mu m^2$. The sensor is mounted on a home-made scanned micro-Hall microscope that allows for high resolution position control (~1μm) above the sample [11]. The BSCCO crystal of size approx. $1 \times 2 mm^2$ was grown by floating-zone technique [12] and its superconducting transition is observed by magnetisation measurements with temperature in the range between 83K and 88K. For magnetisation measurements the Hall voltage is measured using a lock-in amplifier referenced to the drive ac current of 5mA at a frequency of 1kHz. The sensitivity in field is 10μV/G/mA with a noise threshold of the order of 1mG/Hz$^{1/2}$.

The ultra sensitive second-harmonic detection technique that we use to probe the JV-PV interactions is based on the dragging coupling between JVs and PVs. The experiment that is at the basis of the measurement method, and by which JVs drag PVs towards the centre of the sample, is called lensing. This effect is characterised by the symmetry of the magnetisation curve with respect to a magnetic field $H_{ab}$ aligned parallel to the *ab* planes of the crystal (see inset in Fig.1 in [13]). When increasing $H_{ab}$ from zero to a maximum field at constant rate, JVs are created and drag pre-existing PVs across the sample to the sample centre and increasing the PV density. The key signature of lensing is a V-shaped local induction vs field loop. Depending on temperature, magnetic field, and intrinsic sample parameters like pinning, the V-shape can extend over a large range in $H_{ab}$ field or the local induction can saturate at small $H_{ab}$ field [5, 13]. As mentioned above it is the symmetry of the curve (with respect to the sign of $H_{ab}$) that suggests the ac detection method described below.

Second harmonic measurements

Unlike the lensing experiment described above, there is no *dc* in-plane field involved in this "shaking" technique ($H_{ab}$=0). The sample is excited with a superimposed *ac* magnetic field $H^{AC}_{ab}$ produced by a set of Helmholtz coils and aligned with the sample plane. $H^{AC}_{ab}$ is exciting the JV lattice and driving JVs through the sample close to their equilibrium position as illustrated in figure 1(a). Since the dragging interactions are symmetric with respect to the sign of the *ab* plane field, i.e. the local induction is symmetric with respect to the polarity of $H_{ab}$ in the lensing experiment [13], the dragging forces do not depend on the vorticity of JVs. It then follows that the driving force is similar to an ac excitation field at twice the frequency (*2f*) with a phase lag of 90°, as illustrated in figure 1(b) [14]. The Hall sensor is driven by a dc current of 7mA and the lock-in amplifier, used to measure the Hall voltage, is now referenced to the excitation ac field $H^{AC}_{ab}$ and is locked on the second harmonic (*2f*) response $B''$. We use the quadrature component at *2f*, $B''_2$, to reveal JV-PV interactions. Thus, the measurement technique provides two parameters to probe the vortex interaction; (i) the amplitude of the ac excitation field $H^{AC}_{ab}$ and (ii) the frequency *f* of $H^{AC}_{ab}$. Since the residual pick-up associated with the first harmonic is rejected, the noise threshold is reduced to the order of 10μG/Hz$^{1/2}$.

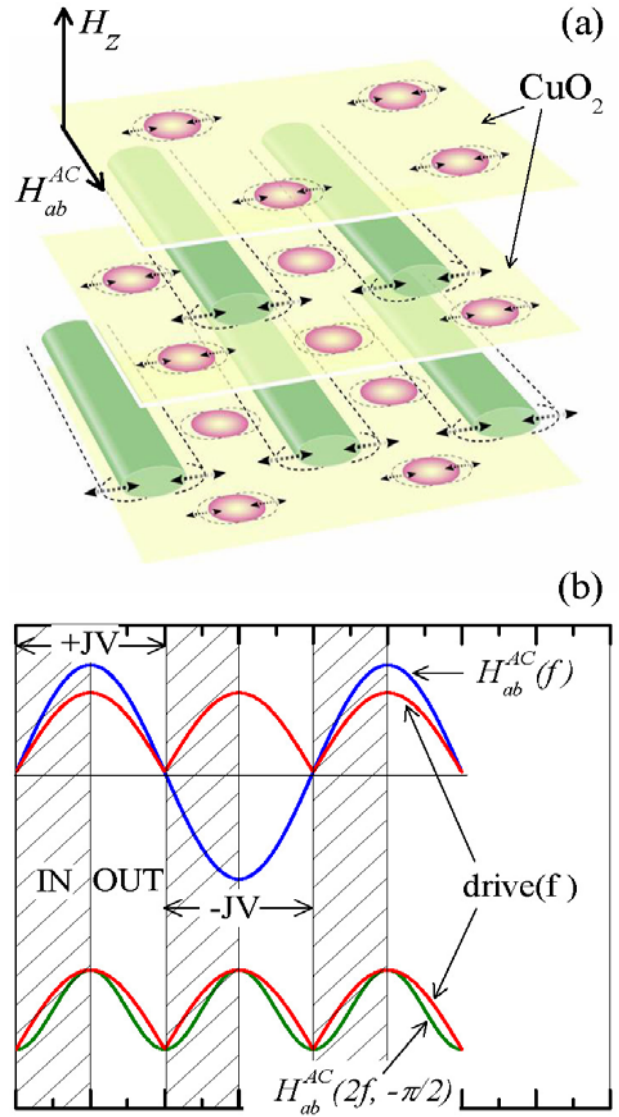

Figure 1. Schematics of the measurement method.
(a) An external ac magnetic field $H^{AC}_{ab}$ is applied parallel to the crystal superconducting $CuO_2$ planes and induces shaking JVs (elongated green ellipsoids in between $CuO_2$ planes), which in turns vibrate the PV lattice created by fixed perpendicular field $H_Z$ (pink pancakes) via the dragging interaction.
(b) The ac magnetic field $H^{AC}_{ab}$ (blue curve) drives the JVs with frequency *f*. In dashed areas, with increasing driving force, Josephson vortices (+JV) and antivortices (-JV) are pushed into the sample, whereas in blank areas where the driving force is decreasing +/-JVs are removed from the sample. The dragging interaction does not depend on the vorticity of the JVs but on their direction of motion, hence dashed areas (respectively blank areas) are equivalent from the point of view of the driving force on the PVs. Thus, the driving force can be represented by an absolute function of $H^{AC}_{ab}$ (red curve), which is similar to an ac drive with frequency *2f* and phase lag of 90° (green curve). Consequently the dragging interaction can be revealed by the quadrature component of the Hall voltage locked on the second harmonic (*2f*) response.

## III. Results

The sensitivity of the ac technique was previously demonstrated [13] by showing that the minimum amplitude of $H^{AC}_{ab}$ to drive PVs with JVs, forming interacting PV and JV lattices, is slightly less than 1Oe (with $H_Z$=8Oe) as shown in the inset to figure 2(a) and is consistent with Hall probe microscopy experiments [10]. The experimental set up has been improved, as well as the signal-to-noise ratio, so to increase the maximum amplitude of the excitation ac field $H^{AC}_{ab}$ up to 14Oe. Figure 2(a) shows $B''_2$ measured above and near the centre of the sample with amplitude of $H^{AC}_{ab}$ at a frequency $f$=120Hz and at a temperature $T$=70K.

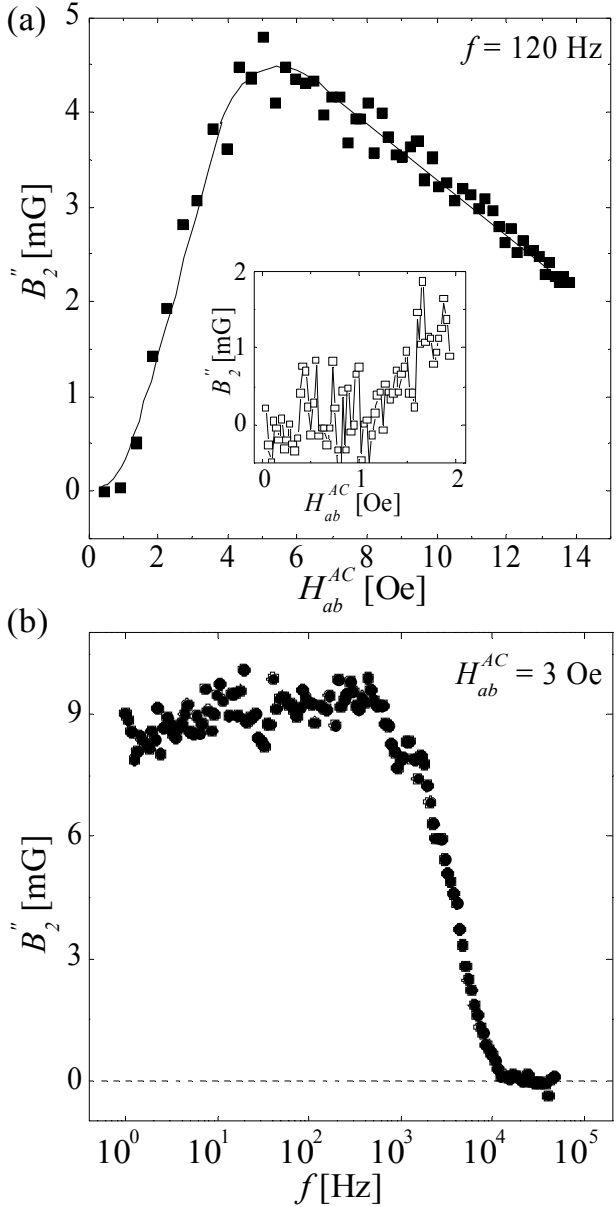

Figure 2. Quadrature component of the second harmonic signal $B''_2$ measured at $T$=70K with $H_Z$=8Oe. (a) Above the central area of the sample as a function of in-plane ac field amplitude at frequency $f$=120Hz. Inset: Low field dependence of $B''_2$ measured under the same conditions [13]. (b) Near the central area of the sample as a function of in-plane ac field frequency with rms amplitude $H^{AC}_{ab}$=3Oe.

The JV-PV interaction gets stronger with the ac amplitude increasing up to $H^{AC}_{ab}$~5Oe and is then reduced above 7Oe with a perpendicular field $H_Z$=8Oe. In other words, the strength of the JV-PV interaction is weakened when the effective applied field is tilted more than 45° with respect to a vertical axis. Figure 2(b) shows the frequency dependence of $B''_2$ measured above the central part of the sample at fixed rms amplitude $H^{AC}_{ab}$=3Oe. The flat background electronics response, which is subtracted from the sample response, was measured under the same conditions far above the sample at sample surface height h=1.5mm where the sample-induced magnetic screening is less than 1%. The JV-PV interaction drops off at frequencies above a few kHz and vanishes above 10kHz. Huge improvement in the signal-to-noise ratio has been achieved since the first frequency dependence published in [13] and is mainly due to an InSb sensor with higher electronic mobility. Both measurements shown in figure 2(a) and 2(b) were performed near the centre of the sample. Nevertheless the magnitude of the response is quite different when comparing the two measurements at the same parameters ($H^{AC}_{ab}$=3Oe and $f$=120Hz). The change in magnitude when slightly changing the position of the sensor above the sample is not negligible and stimulated us to investigate the position dependence of the response, revealing a coupling between the two vortex species which strongly depends on the position in the sample. In order to investigate a possible variation in the vortex coupling across the sample we measured the second harmonic response in temperature at various positions above the sample. Figure 3(a) shows the temperature dependence of $B''_2$ measured with $H^{AC}_{ab}$=3Oe and frequency $f$=120Hz at two different positions above the sample. In addition to the striking change of sign when crossing the sample, the response is characterised by two distinct features in agreement with previous observations [13]; a high temperature response close to Tc, which was referred to as the "peak response" in [13] and a broader response ("bump") at lower temperature. It is the coldest feature, the "bump" indicated with arrows in figure 3(a), that reveals the strength of the interaction between JV and PVs [13]. Figure 3(b) shows the local induction $B$-$H_Z$ measured across the sample cooled in zero field down to $T=30K$ where a field $H_Z=10Oe$ was then applied. On the same graphic we use the maximum of the bump feature, $B''_B$, to compare the JV-PV coupling strength with the position across the sample. $B''_B$ (right hand scale) is extracted from a collection of 11 temperature sweeps recorded approx. every 150μm along the slice where the local induction was measured (left hand scale).

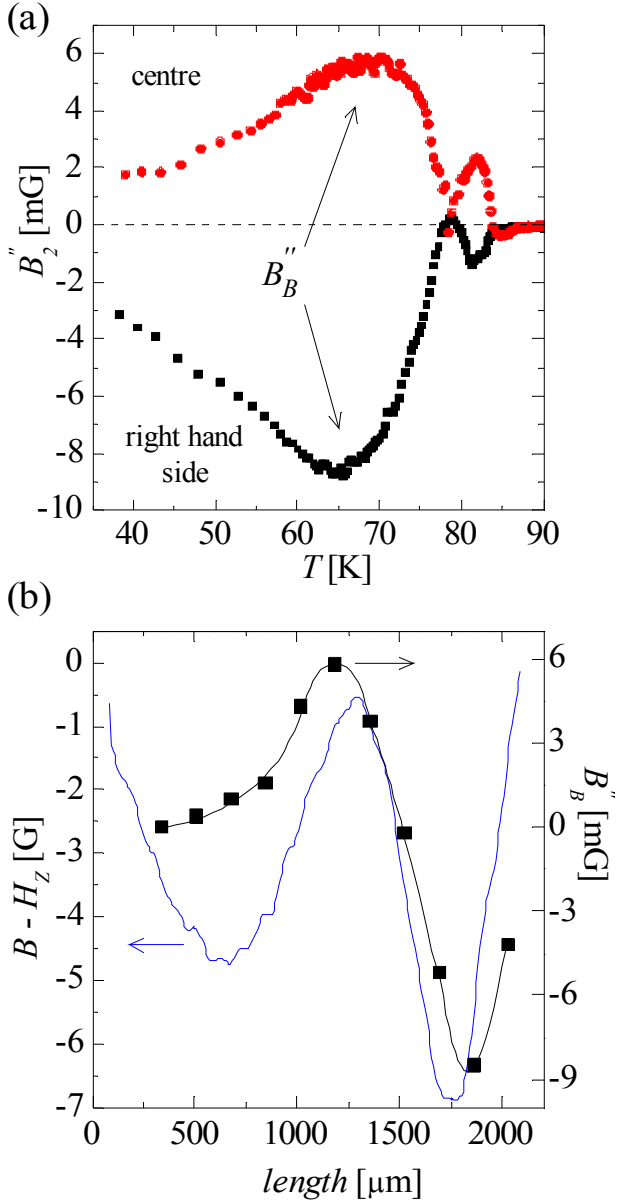

Figure 3. Variation of coupling strength across the sample in field $H_Z=10Oe$ and parameters $H^{AC}_{ab}=3Oe$ and $f=120Hz$.
(a) Temperature dependence of $B''_2$ measured above the central part of the sample (red circles) and above the right-hand side (black squares). Arrows indicate the maximum in the "bump" feature, the magnitude of which $B''_B$ is plotted in the lower figure (b).

(b) Left hand scale: profile of the local induction measured across the sample (blue line) after zero field cooling and $H_Z=10Oe$ applied at $T=30K$. Right hand scale: maximum of the "bump" feature $B''_B$ extracted from temperature sweeps (top figure) at a few positions close to the slice where the profile was measured.

The response is asymmetric across the sample, with respect to the central area, and is correlated to the profile of the local induction, reaching a maximum interaction strength where the field almost completely penetrates the sample in the central area ($B \leq H_Z$), and on the right hand side where the applied field $H_Z=+10G$ is almost completely screened ($B-H_Z=-7G$).

IV. Discussion

Grigorenko and co-workers tentatively depicted the phase diagram for BSCCO single crystals under tilted magnetic fields [10]. They established the boundaries for the vortex states as observed by Hall probe microscopy. While the applied magnetic field is tilted away from the high-symmetry crystalline axis perpendicular to the sample superconducting planes, the vortex state experiences many changes from the very ordered Abrikosov pancake lattice without JVs to the vortex chain state where PVs are well aligned by JVs when the field is parallel to the *ab* planes. As pointed out in [4], the possibility to develop flux devices depends crucially on the existence of the crossing PV and JV lattices where the interaction is expected to be the most useful for applications based on the dragging interaction.

Indeed, three different regimes are observed when sweeping the amplitude of the in-plane shaking field, as illustrated in figure 2(a), while the perpendicular field is fixed. The dragging interaction, hence a second harmonic response, exists only above some minimum *ab* field once the crossing lattice is formed. Within this regime the response is linear with respect to the external perturbation, suggesting a coherent motion of the vortices. Once the response has reached some maximum, which at the measured temperature and field is achieved for $H^{AC}_{ab}\sim 6Oe$, the system enters another linear regime where the strength of the interaction is now lowered while further increasing the excitation. A quantitative comparison with the experimental $H_Z$-$H_{ab}$ phase diagram proposed in [10] suggests that the system is experiencing a crossover from the crossing lattice to the composite lattice. In the latter case PVs tend to align with JVs. Simply speaking JVs oscillate between two extreme positions according to the oscillating (sinusoidal) excitation field ($H^{AC}_{ab}$) and the displacement of the PVs is restricted within the oscillating motion of the JVs close to their equilibrium position. In other words, when PVs start to form chains aligned with JVs, the dragging interaction is reduced because PVs are effectively pinned close to JVs equilibrium positions. The crossing lattice is then the most favourable state for applications based on lensing or ratcheting [4].

In the experiment presented here PVs are vibrating, and the frequency is a probing parameter which is not affecting the physics of the system. Whereas in lensing or ratchet experiments PVs are being moved macroscopic distances across the sample and their motion can be affected by pinning, thus modifying the response depending on the quality of the sample. In principle at low frequency (and high enough $H^{AC}_{ab}$ amplitude) PVs are more sensitive to pinning centres since they can potentially experience a larger displacement trough the sample, resulting from a larger period of the excitation field. The second harmonic response was measured at low field amplitude within 5 decades of frequency above 1Hz (see figure 2(b)), and reveals an interaction which is quasi-frequency independent below 300Hz, with only a tiny increase with frequency within 3 decades. The PV-JV interaction drops off above a kHz and the high frequency limit above which the JVs move to fast to drag PVs is clearly reached at 10kHz. These frequencies are limiting parameters for lensing and ratchet experiments.

The temperature dependence of the second harmonic response shows two features; a *peak* close to, but below, Tc and a broader *bump* feature at lower temperature. The peak reveals the melting of the (pancake) vortex lattice and the bump is a measure of the strength of the JV-PV coupling with the parameter space mapped out [13].

These new results show a change of sign in the second harmonic response when crossing the sample. There are two known possible origins to explain this change of sign. One has to keep in mind that the second harmonic measurements presented here are intimately related to the lensing experiment, because the sign of $B''$ is related to the shape of the lensing curve close to zero field. Thus a change of sign in $B''$ can reveal a change from lensing to anti-lensing. Indeed it has been anticipated [5], assuming conservation of the total number of vortices in the sample (at least at low applied fields), that while vortices accumulate at the centre of the sample the same number of vortices are removed at the edges. However, this behaviour is expected to be symmetric with respect to the sample geometry, hence $B''_2$ across the sample should be symmetric with respect to the centre. Surprisingly, figure 3(b) shows an asymmetric response across the sample which then cannot be explained in terms of competition between lensing and anti-lensing. The second possible origin of the change of sign in $B''$, still somehow related to lensing effect, is a change in the polarity of the local magnetic field. In other words, the local field produces anti-PVs instead of PVs. Indeed figure 3(b) shows that the change of sign in the $B''_2$ curves between the two extrema takes place while the polarity of the local field is reversed, i.e. from a very weak screening central area where the local screening field H≥0 to a strong screening area on the right hand side where H≈-$H_Z$. Clearly field penetration is revealed by the second harmonic measurements.

Further investigation of the dynamics of the compound vortex system can be achieved by measuring the second harmonic response as a function of the amplitude (and the frequency) of the excitation ac field $H^{AC}_{ab}$, and expanding the range of temperature and perpendicular magnetic field. In particular the response vs amplitude of $H^{AC}_{ab}$, as shown in figure 2(a), is very likely to exhibit interesting features such as temperature and field dependent maxima, providing a precise determination of the boundaries between vortex states in the phase diagram of BSCCO single crystals anticipated by Hall probe microscopy experiments [10]. In addition, the technique allows to probe the response at different time scales ~1/*f*.

V. Conclusions

We show how to probe the coupling between Josephson and pancake vortices that coexist in anisotropic superconductors by shaking the Josephson lattice with an ac magnetic field and measuring the second harmonic of the Hall signal associated with the pancake lattice. Variations in the experimental parameters (probing amplitude and frequency) allow for investigations of the states in the vortex matter. The response is shown to be consistent with the profile of the local induction measured independently above the sample, thus providing a test for the validity of the measurement method.

Within our simple approach, the probing method predicts a phase lag of 90° between the field exciting the JVs and the response associated with dragged PVs. This naïve prediction does not take into account any variations of the phase in the Hall signal due to the particular dynamics of the considered vortex system when changing the temperature and the magnetic field.

Interestingly this experimental approach could be readily extended to other types of binary mixtures.

The work was supported by Leverhulme grant F/07 058/V.